\documentclass[11pt]{article}
\usepackage[utf8]{inputenc}
\usepackage[margin=1in]{geometry}
\usepackage{setspace}
\usepackage{amsmath,amssymb}
\usepackage{enumitem}
\usepackage{graphicx}
\usepackage{natbib}
\usepackage{hyperref}
\hypersetup{hidelinks}
\usepackage{bm}
\usepackage{titling}
\usepackage{blindtext}
\usepackage[table,xcdraw]{xcolor}
\usepackage{subfig}

\newcommand{\Real}{\mathbb{R}}

\newcommand{\Ybar}{\bar{Y}}
\newcommand{\tilR}{\tilde{R}}
\newcommand{\Var}{\textrm{Var}}
\newcommand{\fpr}[1]{\left(#1\right)}
\newcommand{\tpr}[1]{\left[#1\right]}
\newcommand{\V}[1]{{\bm{\mathbf{\MakeLowercase{#1}}}}} 
\newcommand{\M}[1]{{\bm{\mathbf{\MakeUppercase{#1}}}}}

\newcommand{\rev}[1]{#1}

\title{\rev{Statistical Methods for Crossover Trials: A Review of Classical and Recent Developments} \vspace{-2ex}}
\author{Salil Koner\thanks{Assistant Professor of Statistics, University of California Riverside, Riverside, CA, USA. Email: \href{mailto:salil.koner@ucr.edu}{salil.koner@ucr.edu}} \\[0.5ex]
{\normalsize Department of Statistics, University of California, Riverside, CA, USA}}
\date{September 30, 2026}
\begin{document}
\maketitle
\begin{abstract}
A comprehensive review of the literature on crossover design is needed to highlight its evolution, applications, and methodological advancements across various fields. Given its widespread use in clinical trials and other research domains, understanding this design's challenges, assumptions, and innovations is essential for optimizing its implementation and ensuring accurate, unbiased results. This article extensively reviews the history and statistical inference methods for crossover designs. A primary focus is given to the AB-BA design as it is the most widely used design in literature. Extension from two periods to higher-order designs is discussed, and a general inference procedure for continuous response is studied. Analysis of multivariate and categorical responses is also reviewed in this context. \rev{Recent developments, including causal (potential outcomes) formulations of carryover, efficient use of period-specific baselines, generalized estimating equations for repeated measurements within periods, and estimands for incomplete crossover trials, are also discussed. Several} open problems in this area are shortlisted.
\end{abstract}

\section{Introduction} \label{intro}
Experimental design is essential to systematically assess the relationship between factors and responses. In a \textsl{Crossover} design, each experimental unit receives more than one treatment during different periods. As the treatment applied to a specific unit changes over periods, it is often called \textsl{change-over} or \textsl{switch-over} design. As per \cite{jones2014design}, the earliest use of crossover design was in agricultural experiments around the mid-nineteenth century by \cite{liebig1847researches}. \rev{Over} the last two decades, crossover designs have been extensively used in phase-1 or phase-2 trials of pharmaceutical studies \citep{lui2016crossover}. \rev{However, their use is not limited to clinical trials; they are also used} in biological assay \citep{finney1956cross}, for evaluation of food products by FDA \citep{brown1980crossover}, weather modification experiments \citep{moran1959power}, psychological studies \citep{ruiz2007framework}, bio-equivalence studies \citep{jaki2013estimation}, \rev{and} consumer preference experiments \citep{harker2008eating}. \cite{senn2002cross} applies such designs in pharmacokinetic studies. Several books on crossover designs, such as \cite{jones2014design}, \cite{ratkowsky1992cross}, offer a detailed description of the experiments where crossover designs have been used historically. \rev{More recently, the crossover principle has been extended beyond individually randomized trials. In cluster randomized crossover trials, entire clusters (e.g., hospitals or intensive care units) switch between interventions over periods \citep{arnup2017understanding, hemming2020use}, and dedicated sample size methods that account for within-period and between-period correlations have been developed \citep{kasza2019impact, li2019power}. Series of N-of-1 trials, in which a single patient repeatedly crosses over between treatments, are another extension whose aggregated analysis borrows heavily from the crossover literature \citep{araujo2016understanding}.}

The extensive use of crossover design stems from a major drawback of traditional parallel group design, where each subject is randomly assigned to one of two or more treatment groups. To illustrate this, assume we are interested in comparing treatments A and B. In a parallel group design, the experimental units chosen for the study are randomly assigned to one of two treatment groups, where one group receives treatment A and the other receives treatment B. To specify distinctly, no subject receives more than one treatment in a parallel group design. Let us call this design $d^\ast$. To test for treatment contrast, we consider the traditional model,
\setlength{\belowdisplayskip}{9pt} \setlength{\belowdisplayshortskip}{2pt}
\setlength{\abovedisplayskip}{9pt} \setlength{\abovedisplayshortskip}{2pt}
\begin{align}
   Y_{ik} = \mu + \tau_{d^\ast(i)} + E_{ik} \quad \quad i=1,2 \quad k=1,2, \;\dots, \; n_i \label{par_mod}
\end{align}
where $Y_{ik}$ is the response for the $k$th subject in $i$th treatment group, $\mu$ be the general mean, $\tau_{d^\ast(i)}$ be the (fixed) effect of treatment applied to the units in group $i$ under the design $d^\ast$ and $E_{ik}$ be the error component. The number of experimental units in the $i$th group is $n_i$, $i=1,2$. This implies that under the above design $d^\ast$, $d^\ast(1)=A$ and $d^\ast(2) = B$. It is very natural to split the error into two components, i.e., $E_{ik} = S_{ik} + e_{ik}$, where $S_{ik}$ be the (random) effect due to $k$th subject in $i$th group and $e_{ik}$ be the random measurements or experimental error. Also, $e_{ik}$ and $S_{ik}$ are independently distributed with $N(0, \sigma^2)$ and $N(0, \sigma_s^2)$ respectively. Although these two errors are not identifiable under this setup, nothing prohibits us from conceptually splitting the total random component in the above way. The idea is whether we can differentiate this random subject effect from the estimators of treatment contrast by considering some other design if the above additive split of $E_{ik}$ is correct. The least squares estimator for the treatment effect $\tau_A - \tau_B$ is $\bar{Y}_1-\bar{Y}_2$ (where $\bar{Y}_i$ is the group mean $i=1,2$) with its standard error (SE) (assuming $n_1=n_2=n$) as $\sqrt{\frac{2(\sigma_s^2+\sigma^2)}{n}}$. In most experimental designs, the inherent heterogeneity in the experimental units induces a high between-subject variability in the response. Then, the SE will be inflated if $\sigma_s^2$ is large compared to $\sigma^2$. Naturally, an analysis of variance (ANOVA) type test for detecting a significant treatment difference has low power. On the other hand, in a crossover design where each subject receives more than one treatment in different periods, it is possible to estimate every treatment contrast (and possibly contrasts of all other effects) through within-subject difference after eliminating variability due to the subject. If the magnitude of within-subject variation $\sigma^2$ is minimal, then the test can be done with high power. It is reasonable to contemplate that there can be a confounding effect of the period on the response, but this can be eliminated by splitting the units into two groups where one group receives A in period 1 and B in period 2\rev{, and the other} receives B in period 1 and A in period 2. Such a design, \rev{known} as $2 \times 2 \times 2$  design or AB-BA design, is the most commonly used crossover design in the literature, compared to the other complex crossover designs where the objective is to compare more than two treatments and/or responses are observed for each subject over more than two periods. Also, when we have more than two treatments, the models for the carryover effect become more complex. We will discuss those in more detail in Section~\ref{higherorder}. Moreover, crossover designs are generally more economical as fewer subjects are required to get the same number of responses compared to parallel-group studies.

One potential objection to this crossover design is that the effect of treatment in period 2 is confounded by the residual effect of the treatments applied in period 1. This residual effect is termed as \textsl{carryover} effect of treatment \citep{cochran1941double}. In most trials, the objective is to measure the direct impact of the treatment on the response after adjusting for carryover. Efficient estimation of direct treatment effect through the within-subject difference in the presence of significant carryover can be done by choosing a suitable crossover design with more than two periods and/or more than two groups. However, carryover hinders a robust estimation of the direct treatment effect in an AB-BA design. Much research has been done in the late twentieth century to address this problem. We will discuss those in Section~\ref{AB-BA}. \rev{Crossover designs are not appropriate in applications where the treatment substantially changes the initial condition of the experimental unit.} For example, if a surgery completely cures a patient, the patient's condition before and after surgery is not the same. With that perspective, crossover designs are particularly effective in assessing treatments that can't cure the subject entirely but moderate the disease, such as migraine or asthma \citep{senn2002cross}.

The rest of the paper is organized as follows. Section \ref{notation} consolidates the list of notations we will use throughout the rest of the paper. In Section~\ref{AB-BA}, we study the existing inference procedures for AB-BA design when the response is continuous. Further, we discuss the advantages of higher order designs in Section~\ref{higherorder} and review some inference procedures. In Section~\ref{baseline}, we revisit AB-BA design to explore the use of baseline measurements. Section~\ref{multivariate} includes the inference techniques for a multivariate response. Statistical methods for analyzing binary or categorical responses have been discussed in Section~\ref{binary}. Concluding remarks and possible areas of further research are summarized in Section~\ref{open}.
\vspace{-0.2cm}
\section{Notation} \label{notation}
We will stick to the following notation and terminology for the rest of the paper. \rev{A crossover design is defined by three parameters:} $t$ : number of treatments to be compared, $g$ : number of treatment groups, $p$ : number of periods over which responses are observed. Throughout the paper, we denote a crossover design with these parameters as $d(t,p,g)$. It is crucial to notice that, unlike parallel group design, the number of treatment groups ($g$) is not always equal to the number of treatments to be compared ($t$). Here, a treatment group is defined by a sequence of treatments, and the sequence length is the number of periods in the study. The experimental units selected for the study are randomly assigned to one of the $g$ treatment groups. We define $n_i$ as the number of units in the $i$th group, $i=1,2,\;\dots,\; g$. $n_1 + n_2 + \dots + n_g = N$, the total number of subjects or units in the study.

 Consider a small example with $t=2$, $p=3$ and $g=4$. There are two treatments to be compared (say A and B), each subject is assigned to one of the four treatment groups, and the response is observed over three periods. An example of a $d(2,3,4)$ design is presented in \rev{Table} \ref{tab:crossdesign}(a). Each row defines a treatment group, and the columns represent the period. The four treatment groups are ABA, BAB, AAB, and BBA. Each cell represents the treatment applied to a subject in a treatment group and a particular period. For example, a subject in treatment group $3$ receives treatment A at period $1$, again A in period $2$ and B in period $3$. This implies that each subject gets multiple treatments throughout the study. In other words, the treatments are switched over across the periods; this is why it is called a crossover or switch-over design.  Similarly, \rev{Table} \ref{tab:crossdesign}(b) provides a diagram of another crossover design with three treatments (A, B, and C), four periods, and six treatment groups, i.e., a $d(3,4,6)$ design. To clarify, a subject in group $5$ receives treatment B in period $1$, A in period $2$, and treatment C in periods $3$ and $4$. Subjects in group $3$ start with treatment C, followed by A, B, and B again over the four study periods.


\begin{table}[h!]
\centering
\subfloat[Diagram of a d(2,3,4) design]{
\begin{tabular}{|c|lll|}
\hline
\multicolumn{1}{|c|}{{\color[HTML]{3531FF} Group}} & \multicolumn{3}{c|}{{\color[HTML]{3531FF} Period}} \\ \cline{2-4}
\multicolumn{1}{|c|}{} & 1 & {\color[HTML]{000000} 2} & 3  \\ \hline
1 & A & B & A \\
2 & B & A & B \\
3 & A & A & B \\
4 & B & B & A \\ \hline
\end{tabular}}
\qquad
    \subfloat[Diagram of a d(3,4,6) design]{\begin{tabular}{|c|llll|}
\hline
\multicolumn{1}{|c|}{{\color[HTML]{3531FF} Group}} & \multicolumn{4}{c|}{{\color[HTML]{3531FF} Period}} \\ \cline{2-5}
\multicolumn{1}{|c|}{} & 1 & {\color[HTML]{000000} 2} & 3 & 4 \\ \hline
1 & A & B & C & C \\
2 & B & C & A & A \\
3 & C & A & B & B \\
4 & A & C & B & B \\
5 & B & A & C & C \\
6 & C & B & A & A \\ \hline
\end{tabular}
    }
\caption{\rev{Two examples of crossover designs: (a) a $d(2,3,4)$ design with two treatments, three periods, and four treatment groups; (b) a $d(3,4,6)$ design with three treatments, four periods, and six treatment groups.}}
\label{tab:crossdesign}
\end{table}

$Y_{ijk}$ is the single variate response for $k$th subject in $i$th sequence at period $j$. Except for Section~\ref{binary}, the response is assumed to be continuous. The parameter $\mu$ is termed as the overall/general mean, $\pi_j$ as the effect of $j$th period, $\tau_{d(i,j)}$ as the direct effect of treatment administered to the $i$th group at period $j$ and $\lambda_{d(i,j-1)}$ as the carryover effect of treatment at period $j-1$ for $i$th group. For example, under the design $d(3,4,6)$ in \rev{Table} \ref{tab:crossdesign}(b), $d(2,3)=A$, $d(4,2) = C$, $d(5,1) = B$ etc. It is assumed that there is no carryover effect in the first period, i.e., $\lambda_{d(i,0)}=0$.  $S_{ik}$ be the effect of $k$th subject in group $i$. Unless otherwise specified,  $S_{ik}$ is assumed to be random.  Lastly, $e_{ijk}$ is the random error of $Y_{ijk}$. We refer the subscript $i$ as the sequence, $j$ as the period, $k$ as the subject ($i=1\dots g, j=1\dots p, k=1(1)n_i$ and $\sum_{i=1}^g n_i = N$). Now, we will define certain terms that are general to any crossover design but are mainly introduced for AB-BA design.
\begin{align}
    &T_{ik} = Y_{i1k} + Y_{i2k} \;\; ; \quad P_{ik} = Y_{i1k} - Y_{i2k} \;\; ;\quad C_{ik} = \begin{cases} Y_{i1k} - Y_{i2k} \; &\textrm{if } i=1 (\textrm{ group AB}) \\ Y_{i2k} - Y_{i1k} \; &\textrm{if } i=2 (\textrm{ group BA})
    \end{cases} \label{eqn: notation}
\end{align}
$T_{ik}$, $P_{ik}$, $C_{ik}$ are called subject totals, period difference, and crossover difference, respectively. $\bar{T}_{i\bullet}$,  $\bar{P}_{i\bullet}$,  $\bar{C}_{i\bullet}$ be average of them over $k$. An $M \times 1$ multivariate response recorded for $k$th subject of $i$th group in $j$th period is denoted by $\M{Y}_{ijk}$.
\section{Inference procedures for continuous response under AB-BA design } \label{AB-BA}
This section discusses existing methodologies for analyzing continuous response under AB-BA design. The key point is that the response of a subject in group AB at period 2 will depend not only on the effect of treatment B but also on the residual effect of A that is applied in period 1 and similarly for group BA. A traditional model that adjusts for the carryover effect follows.
\begin{align}
    Y_{ijk} = \mu + \pi_j + \tau_{d(i,j)} + \lambda_{d(i,j-1)} + S_{ik} + e_{ijk} \label{trad_mod}
\end{align}
We assume $S_{ik}, e_{ijk}$ are independently distributed with mean $0$, variance $\sigma_s^2$ and $\sigma^2$ respectively. Under model~(\ref{trad_mod}), the expected value of $Y_{ijk}$ for four different cells are presented in \rev{Table} \ref{tab:expres}.
\begin{table}[h]
\centering
\begin{tabular}{|c|c|c|c|c|}
     \hline
     Sequence & Period-1 & Period-2 & Period Diff & Period Total \\
     \hline
     $E(Y_{1jk})$ & $\mu+\pi_1+\tau_A$ & $\mu+\pi_2+\tau_B+\lambda_A$ & $\pi + \tau - \lambda_A$  &  $2\mu+\pi_1+\pi_2+ \tau_A+\tau_B+\lambda_A$\\
     \hline
     $E(Y_{2jk})$ & $\mu+\pi_1+\tau_B$ & $\mu+\pi_2+\tau_A+\lambda_B$ & $\pi - \tau - \lambda_B$  & $2\mu+\pi_1+\pi_2+ \tau_A+\tau_B+\lambda_B$ \\
     \hline
     Diff & $\tau$ & $\lambda$ - $\tau$ & $2\tau - \lambda$ & $\lambda$ \\
     \hline
\end{tabular}
\caption{Expected response for different cells in an AB-BA crossover design}
\label{tab:expres}
\end{table}
The \rev{parameters of interest} $\pi=\pi_1-\pi_2$, $\tau=\tau_A-\tau_B$, $\lambda=\lambda_A - \lambda_B$ \rev{are the contrasts} of period, direct treatment and carryover effects respectively. Also, as an identifiability condition, we set $\lambda_A+\lambda_B=0=\tau_A+\tau_B = \pi_1 + \pi_2$. From the table, it is clear that we can't construct an unbiased estimator for the carryover effect based on within-subject differences. Moreover, under Gaussian assumption, assuming $n_1=n_2=n$
\begin{align}
    &T_{\tau,W} = \left(\bar{P}_{1\bullet}-\bar{P}_{2\bullet}\right)/2 \sim N\left(\tau - \lambda/2, \; \sigma^2/n\right) \;\;\; \textrm{and} \quad \;\;   T_{\tau,B} = \left(\Ybar_{11\bullet}-\Ybar_{21\bullet}\right)  \sim N\fpr{\tau, \; 2(\sigma_s^2+\sigma^2)/n} \label{eqn:treatEst} \\
    &T_{\pi,W} = \left(\bar{C}_{1\bullet}-\bar{C}_{2\bullet}\right)/2 \sim N\left(\pi - \lambda/2, \; \sigma^2/n\right)
    \;\;\; \textrm{and} \quad \;\;   T_\lambda =\bar{T}_{1\bullet}-\bar{T}_{2\bullet} \sim N\left(\lambda, \;4(2\sigma^2_s +\sigma^2)/n\right) \label{eqn:otherEst}
\end{align}
The estimator $T_{\tau,W}$ in~(\ref{eqn:treatEst}) based on within-subject difference is unbiased for treatment effect only if there is no carry-over effect. So, we can do a student-t or F test using this only if $\lambda=0$. An unbiased estimator $T_{\tau,B}$ for the treatment effect ($\tau$) based on only first-period responses as in~\rev{(\ref{eqn:treatEst})} can be constructed. However, that has a higher standard error due to between-subject variability. The criticism is that if we estimate $\tau$ based on only one period, then \rev{the rationale for conducting a crossover study is lost}. However, this is exclusively a problem for AB-BA crossover design. We will see in the next section that this problem can be alleviated by studying for more than two periods. Many researchers recommended incorporating washout periods between the two periods of the study to nullify the effect of carryover \citep{senn1996ab}. However, including a washout period is not always feasible because it will elongate the duration of the study or due to ethical reasons \citep{bose2009optimal}. \rev{Moreover, a washout period removes only the biological (pharmacological) carryover. In comparative effectiveness research, the carryover can be \textsl{behavioral}, i.e., the treatment received in period 1 alters the participant's behavior (e.g., adherence) in period 2, and such carryover cannot be eliminated by a washout period \citep{shi2024behavioral}.} Another issue of the AB-BA design is a treatment-by-period interaction and a carryover effect that might affect the response. In a $2 \times 2 \times 2$ crossover study, the carryover effect is aliased with this interaction effect \citep{armitage1982two}. One can understand this by noting that there are only 4 cell means, which means a total of 3 degrees of freedom (d.f) can be accounted for, out of which 2 of them are attributed to treatment and period effect. So, only one d.f is left to identify any other effect. Still, many researchers have worked on developing methodologies for testing treatment and carryover effects in an AB-BA crossover design over the past few decades because 1) it is very easy to implement the randomization of subjects to groups and treatments compared to more complicated design, 2) If a design lasts longer than two periods, then subject drop-out can pose difficulty in the analysis \citep{sheehe1961latin}. 3) The cost of recruiting subjects increases as the number of periods increases.
\subsection{Parametric approach for testing fixed effects} \label{AB-BAPar}
\cite{cox1958planning} analyzed the crossover design, assuming that there is no carryover effect, and pointed out that it is nothing but a split-plot design. \cite{grizzle1965two} pioneered testing for treatment effect in the presence of carryover assuming model~(\ref{trad_mod}) and calculated the analysis of variance (ANOVA) table. A correct table version can be found in the appendix of \cite{hills1979two}. The author concluded that if there is a significant carryover, it is better to use a randomized parallel-group trial than a crossover design. But, when the presence of residual is in doubt, he proposed a two-stage (TS) procedure that first tests for carryover effect based on $T_\lambda$ in~(\ref{eqn:otherEst}) and then tests for treatment effect using the statistic $T_{\tau,W}$ if the test for carryover is not significant, otherwise tests using $T_{\tau,B}$. It is worth noting that the preliminary carryover test lacks power as the standard error of the statistic $T_\lambda$ is influenced by between-subject variation. Hence, the author recommends testing at a higher significance level ($10\%$). Under the normality assumption, the \rev{test statistics} for testing $H_0: \tau = 0$ \rev{are} given by
\begin{align}
    &F_{co} = nT_{\tau, w}^2/\hat{\sigma}^2 \sim F_{1,2n-2}\left(\phi\left(\left(\tau-\lambda/2\right)^2\right)\right) \label{eqn:CO} \\
    &F_{par} =  nT_{\tau, b}^2/2\tpr{\hat{\sigma}^2+\hat{\sigma}^2_s} \sim F_{1,2n-2}\fpr{\phi\fpr{\tau^2}} \label{eqn:PAR} \\
    &F_{TS} = F_{co}\;I\left(\lvert T_\lambda \rvert \leq c\right) + F_{par}\;I\left(\lvert T_\lambda \rvert > c\right) \label{eqn:TS}
\end{align}

\rev{where} $c$ is an appropriate cutoff value and $\hat{\sigma}, \hat{\sigma_s}$ be the maximum likelihood estimators of $\sigma, \sigma_s$ respectively. The subscripts of the F-statistics in~(\ref{eqn:CO})-(\ref{eqn:TS}) are meant to highlight that the tests are based on crossover (co) and parallel group (par) trials and two-stage procedure (TS). The non-centrality parameters vanish when $\tau-\lambda/2=0$ and $\tau=0$, respectively. Confidence intervals (CI) for the treatment effect can be obtained by inverting the above $F$ tests. This can be done using SAS procedure \verb|proc mixed| \citep{SASmixed} or \verb|glm()| function in R \citep{basepack}\rev{.}

\cite{brown1980crossover} analyzed the cost efficiency of the crossover design by incorporating the sample size that would be required to get the same precision and concluded that if there is insignificant carryover, crossover design is \rev{considerably more economical} than a parallel group design, but \rev{it can be costlier} otherwise.

Under the null hypothesis $\tau=0$, $F_{par}$ has a central F-distribution. If $\lambda \neq 0$, the non-centrality parameter in $F_{co}$ does not vanish, and the test becomes biased. \cite{willan1986carryover} pointed out that if there is no carryover effect without the treatment effect, then under $H_0$ $F_{co}$ has a central F-distribution. Consequently, the test becomes unbiased. But, the problem occurs otherwise. They mentioned that for a placebo-controlled trial, or when a new treatment is compared against a standard treatment, it is expected not to have a carryover when there are no treatment effects. They also differentiated between physical and psychological carryover and asserted that significant carryover can exist without treatment effect for the former but not for the latter. The authors compared the power of the crossover and parallel group tests and derived
\vspace{-0.3 cm}
\begin{align}
    P_{co}(\tau, \lambda) >  P_{par}(\tau, \lambda) \quad \textrm{if} \quad  \left\lvert\lambda/\tau\right\rvert < 2 - \sqrt{2(1-\rho)} \label{eqn:powerCOPAR}
\end{align}
where $P$ is the power function of the crossover and parallel group tests and $\rho=\sigma_s^2/(\sigma_s^2+\sigma^2)$ is the intra-group correlation. This means the power of F-tests in (\ref{eqn:CO}) and (\ref{eqn:PAR}) depends on the value of intra-group correlation and the relative magnitude of carryover and treatment effect. Given $\sigma_s^2 > \sigma^2$, if the carryover effect is small relative to the direct treatment effect, the crossover test of $\tau$ can outperform the parallel group test. Later on, \cite{willan1988using} proposed to use the maximum test-statistic $F_{max} = \max\;(F_{co}, F_{par})$ for testing the treatment effect under the assumption that no differential carryover exists in the absence of treatment effects.

Because the test-statistic $T_\lambda$ and $T_{\tau,B}$ are highly correlated, the TS procedure can be biased even if $\lambda$ is small. \cite{freeman1989performance} examined the performance of $F_{co}$, $F_{par}$ and $F_{TS}$ for testing treatment effect but they allowed the possibility that $\lambda$ can be significant even if $\tau=0$. Simulation results showed that the \rev{actual significance level of the TS procedure exceeds the nominal type-I error rate} ($\alpha$) even when $\lambda=0$\rev{, and the inflation grows} as $\lambda$ deviates from zero. In the absence of carryover, $F_{TS}$ and $F_{co}$ are more powerful than $F_{par}$, but $F_{TS}$ achieves superiority at the expense of increased significance level. \cite{senn1996ab} proposed a correction to the TS procedure to maintain the type-I error rate ($\alpha$). If the type-I error of the carryover test is $\alpha_1$, he proposed to conduct the $F_{par}$ test at a reduced significance level of $\alpha \alpha_1$. This procedure is conservative except when within-subject variability $\sigma^2=0$. \cite{wang1997use} \rev{proposed} another two-stage procedure that controls the \rev{type-I} error rate like the above and offers a detailed comparison between maximum test statistics and two-stage procedures. The conclusion is similar: if the relationship between carryover and treatment is unknown, then the two-stage procedures can be very misleading. But, under the assumption that carryover does not persist in the absence of treatment effect, $\lambda/\tau$ is negative and large in magnitude, two-stage procedures are more robust to the possible overestimation that maximum test-statistic or the crossover test can lead to. \rev{Since the above results rely on large-sample approximations or known variance components, \cite{kabaila2016finite} studied the finite-sample performance of the two-stage analysis when the variance components are estimated from a small sample.}

Another disturbing feature of the TS procedure mentioned by \rev{\cite{senn2002cross}} is if there is a significant effect of sequence (which is aliased with the carryover in AB/BA design), the estimator $T_{\tau, b}$ is not unbiased for $\tau$. The test of treatment effect based on only first-period observations will be biased.

\rev{More recently, \cite{shi2024behavioral} revisited the AB-BA design under a potential outcomes framework that requires only randomization and no modeling or normality assumptions. They showed that the commonly used within-subject estimator remains unbiased for the average treatment effect under a population-level no-carryover assumption. When the carryover effect is positive, the estimator underestimates the treatment effect; this does not inflate the type-I error of one-sided tests but reduces power. When the carryover is negative, the treatment effect is overestimated, and the type-I error is inflated. They also derived the condition under which the crossover design remains more powerful than a parallel-group design in the presence of carryover, which complements the power comparison in (\ref{eqn:powerCOPAR}) under minimal assumptions. Moreover, they showed that adjusting for baseline covariates using a working linear model with treatment-by-covariate interactions yields an estimator whose asymptotic variance is never larger than that of the unadjusted estimator, even when the working model is misspecified. This is in line with the recent regulatory guidance encouraging covariate adjustment in randomized trials \citep{fda2023covariate}.}
\subsection{Bayesian approach}

The performances of all the existing frequentist approaches rely on the information on the significance of carryover.
\cite{selwyn1981bayesian} pioneered the Bayesian analysis for the crossover trials in the \rev{bioequivalence} study by incorporating the uncertainty as an informative prior \rev{on} the carryover effect. Arguing that such information is rarely available in practice, \cite{grieve1985bayesian} assumed a completely uninformative prior for the parameters. The posterior distribution of $\tau$ is centered at the difference of the first-period observation, and the second-period observation comes through the variance. This means if we do not assume that the carryover effect is insignificant, then we have to do our analysis based on first-period observations, a similar point made by \cite{grizzle1965two}. To incorporate a prior belief on the carryover effect, \cite{grieve1985bayesian} proposed a Bayes factor approach that quantifies the model's probability with carryover and no carryover. This is more practical than \rev{the} \cite{selwyn1981bayesian} approach because the variance $\sigma_R^2$ in the prior distribution of $\lambda$ is not available in practice. He mentioned that if we have prior knowledge about carryover, then we should have prior information on the treatment effect too, so why not incorporate that information in the prior of $\tau$? The conclusion is that the inference for the treatment effect is heavily based on our prior belief on the carryover effect, \rev{which is also the conclusion of} all the frequentist approaches.



Often, the experimenter does not have an adequate extent of certainty about the carryover effects. Therefore, a complete specification of the prior distribution becomes difficult. \cite{david2002hierarchical} introduced a hierarchical approach for the Bayesian analysis of two-period crossover design to ensure robustness to prior choice. \cite{li2017objective} proposed an objective Bayesian estimation approach for AB/BA crossover trials. The authors assume a mixture distribution of the model with carryover and no carryover similar to \cite{grieve1985bayesian}, in addition to a \rev{Zellner--Siow} prior distribution \citep{zellner1980posterior} for $\lambda$ when carryover is present. The simulation study shows that the treatment effect's nominal coverage probability is maintained close to $95\%$ level, irrespective of whether $\lambda$ is small or large. In terms of MSE, this method significantly improves \rev{upon} the TS procedure when $\lambda$ is moderate.

\subsection{Non-parametric approach} \label{AB-BANonPAr}
Contemporary to \cite{grizzle1965two}'s parametric inference, \cite{koch1972328} proposed a non-parametric approach for analyzing $2 \times 2 \times 2$ crossover design using Wilcoxon rank sum test or equivalently by Mann-Whitney test. Let $R_{ik}, \tilde{R}_{ik}$ be respectively the rank of subject totals $T_{ik}$ and period differences $P_{ik}$ for the kth subject in the ith group based on the entire sample. Let, $R_i = \sum_k R_{ik}$ and $\tilde{R}_i = \sum_k \tilde{R}_{ik}, i=1,2$. The Wilcoxon rank-sum test-statistic for testing $H_{01}: \lambda=0$ and $H_{02}:\tau=0$ under $\lambda=0$, is given by,
\vspace{-0.1 cm}
\begin{align}
    Z_\lambda = \frac{R_1-E(R_1)}{\sqrt{\Var(R_1)}} \quad ; \quad  Z_\tau = \frac{\tilR_1-E(\tilR_1)}{\sqrt{\Var(\tilR_1)}} \label{eqn:NonparZ}
\end{align}
 \rev{where} the expectation and variance of $R_1$ and $\tilde{R}_1$  under the null hypothesis can be obtained from \cite{mann1947test}. If the sample size is large, under null, the test statistics follow an approximately standard normal distribution. The exact CI for $\tau$ can be obtained using the methods of \cite{hodges1963estimates} by computing the median of all possible response differences in the two groups. SAS procedure \rev{\texttt{proc npar1way}} and R function \verb|wilcox.test()| \rev{offer} CI, an exact and asymptotic significance test. It should be noted that the test of treatment effect is valid only if $\lambda = 0$. Otherwise, the test should be done based on period-1 responses. A similar two-stage procedure can also be adopted here to test for the effects of the treatment. \cite{koch1972328} also proposed non-parametric test for testing $H_{03}: \lambda=0, \tau=0$. Under $H_{03}$, the responses for each subject are an identically distributed bivariate random variable, and the Wilcoxon rank-sum test of location following \citet{chatterjee1964non} can be constructed. Let $\gamma_{ijk}$ be the rank of the kth subject in group $i$ on period $j$ based on the entire sample. Define $\M{\gamma}_{ik}=(\gamma_{i1k}, \gamma_{i2k})^\top$ as the rank vector and $\bar{\V{\gamma}}_i = (\bar{\gamma}_{i1}, \bar{\gamma}_{i2})^\top$, where $\bar{\gamma}_{ij}=\sum_{k=1}^{n_i} \gamma_{ijk}/n_i \; i=1,2$ and $\bar{\V{\gamma}} = \fpr{n_1\bar{\V{\gamma}}_1+n_2\bar{\V{\gamma}}_2}/(n_1+n_2)$\rev{.} Then, the test statistic is given by
\begin{align}
    &W_\tau = \fpr{\bar{\V{\gamma}}_1 - \bar{\V{\gamma}}_2}^\top\M{\Sigma}_\gamma^{-1}\fpr{\bar{\V{\gamma}}_1 - \bar{\V{\gamma}}_2} \sim \chi^2_2 \quad (\textrm{under } H_{03}  \textrm{ asymptotically}) \label{eqn:nonparBIV} \\
    \textrm{where } &\M{\Sigma}_\gamma = \frac{(n_1+n_2)}{n_1n_2(n_1+n_2-1)} \sum_{i=1}^2\sum_{k=1}^{n_i} \fpr{\M{\gamma}_{ik}-\bar{\V{\gamma}}}\fpr{\M{\gamma}_{ik}-\bar{\V{\gamma}}}^\top
\end{align}

The non-parametric tests described in~(\ref{eqn:nonparBIV}) are based on the assumption that the variations of the samples are the same for both groups. \cite{cornell1991nonparametric} proposed two non-parametric tests for equality of the dispersion of the responses in two groups. One is based on the rank correlation coefficient of $T_{ik}$ and \rev{$P_{ik}$}, and the other is based on their concordance. The test is based on the fact that under model~(\ref{trad_mod}), $\textrm{Cov}(T_{ik}, P_{ik}) = \sigma_1^2 - \sigma_2^2, i=1,2$, which vanishes under the null hypothesis.

\cite{tudor1994review} proposed a more general non-parametric approach to test for significant total treatment (combined of direct and carryover) effect. To accomplish that, they tested $H_0: 2\tau - \lambda=0$. The test statistics are based on the period differences of two groups $P_{ik}, i=1,2$, which is given by
\begin{align}
    Q = \frac{\fpr{\bar{P}_{1\bullet} - \bar{P}_{2\bullet}}^2}{S_p^2} \quad ; \quad \textrm{where } S_p^2 = \frac{(n_1+n_2)}{n_1n_2(n_1+n_2-1)} \sum_{i=1}^2\sum_{k=1}^{n_i} (P_{ik} - \bar{P})^2 \label{eqn:TSNonpar}
\end{align}

\rev{where} $\bar{P} = (n_1\bar{P}_{1\bullet}+n_2\bar{P}_{2\bullet})/(n_1 + n_2)$. Under $H_0$, if $n_i$ are large ($>15$), $Q$ follows approximately $\chi^2$ distribution with $1$ degree of freedom. The authors mentioned the above sample size requirement for each group to ensure that the total number of subjects in the study is large enough ($>30$) to hold asymptotic results. As they suggested, the above test has a higher power of detecting any significant total treatment effect because it is entirely based on within-subject differences. If the test is rejected, further tests must be done to detect whether the difference is due to direct treatment and/or carryover effects. In the second stage, a test for carryover based on $Z_\lambda$ needs to be done. If the test is rejected, a test for $\tau$ based on first-period responses can be done; otherwise, a test using $Z_\tau$ as in~(\ref{eqn:NonparZ}) can be carried out. Sometimes, knowledge of the length of the washout period can be incorporated to judge the absence of carryover effects. This involves a three-stage procedure. The authors recommended setting the significance levels for each test before the start of the test to ensure that the nominal significance level is not inflated.

\cite{jung1999multivariate} proposed a rank measure association-based non-parametric test under crossover design applicable to an ordinal response.
Consider a randomized parallel group design with two treatments. Let $\mu_i$ be the mean response for the ith group. The difference $\mu_1 - \mu_2$ represents a measure of disparity between the two groups. In the parametric approach, we model this disparity as $\mu_1 - \mu_2 = E(X_{1k}-X_{2k}) = \tau_A - \tau_B$. For the Wilcoxon rank-sum test, we model the median of differences between the responses through $\tau_A - \tau_B$. Similarly, \rev{Jung and Koch} proposed to model the disparity between the two groups $\psi=P(X_{1k}-X_{2k} > 0)$ non-parametrically by  $\tau=\tau_A - \tau_B$. In an AB-BA crossover design, we have $4$ cell means. So, there will be four different rank measures, $\V{\psi} = (\psi_1, \psi_2, \psi_3, \psi_4)^\top$.

If  $\hat{\V{\psi}}$ be the estimate of $\V{\psi} = (\psi_1, \psi_2, \psi_3, \psi_4)^\top$ and $\V{\beta}=(\tau, \pi, \lambda)^\top$ be our parameter of interest, then, they proposed the model $\hat{\V{f}} = \textrm{logit}(\hat{\V{\psi}}) = \M{X}\V{\beta}$ where the design matrix $\M{X}$ can easily be obtained from the relation between $\V{\psi}$ and $\V{\beta}$. The estimates $\hat{\V{\psi}}_m$ \rev{are} constructed non-parametrically using the U-statistics of the responses related in $m$th comparison $m=1\dots 4$, e.g., for the first comparison, we consider only the responses in the first period. The \rev{authors} also derived the asymptotic multivariate normal distribution of the estimator $\hat{\V{\beta}}$ with a suitably chosen covariance matrix, which can be used to test any hypothesis of the form $H_0: \M{C}\V{\beta}=0$ and to derive approximate $100(1-\alpha)\%$ CI. The performance of this method is not shown in the paper through simulation but is applied to real data, and the results were similar to the standard procedures. For an extensive application of the method, see Chapter $2$ of \cite{jones2014design}.


\section{Higher order crossover designs} \label{higherorder}
Non-estimability of the carryover effect under model~(\ref{trad_mod}) is solely a problem of AB-BA crossover design and undermines its effectiveness.
Complex models can be analyzed with higher power using higher-order crossover designs. By higher order, we mean either the design consists of more than two periods and/or the subjects are randomly assigned to more than two groups. For example, instead of having two groups, we can have four treatment groups: AB, BA, AA, and BB. Sometimes, it is impractical to stretch the study to more than two periods. In those situations, randomizing subjects into more than two sequence groups is essential for efficient estimation of treatment effects. Any design can be specified by three parameters: number of treatments ($t$), number of periods ($p$), and number of groups ($g$). The tests for treatment and carryover effects gain power because each effect in the model~(\ref{trad_mod}) can be estimated using within-subject difference. Moreover, the carryover, treatment by period interaction, and group effects are identifiable and can be estimated separately. The estimability of effects in any crossover design can be understood in the following way. Consider a design $d(t,p,g)$, with $pg$ cell means, which means there are $pg-1$ degrees of freedom, out of \rev{which} $g-1$ \rev{are} attributed to group effects, $p-1$ \rev{to} period effects, \rev{and} the remaining $(p-1)(g-1)$ d.f \rev{are} split \rev{among the} treatment effect, treatment by period interaction, \rev{and} carryover effect.

\subsection{Complex carryover models}

For higher order designs with more than two periods, the model in~(\ref{trad_mod}) assumes that the carryover effect of a treatment in the period $j$ depends only on the treatment in period $j-1$. This is called first-order carryover. Moreover, the carryover effect of a treatment on itself (self-carryover) is assumed to be identical to the carryover of the treatment on another treatment (mixed-carryover), e.g. in the design $d(2,3,4)$ shown in \rev{Table} \ref{tab:crossdesign}(a), although the second group receives treatment A and the group $4$ receives the same treatment (B) again at period $2$, carryover effect of B from the first period in group $2$ \rev{is} assumed to be the same as the carryover of B in group $4$. However, in many practical applications, this simple carryover model is inadequate. For example, in customer survey experiments for the evaluation of certain products, if a person strongly likes or dislikes a product, its lingering effect influences the evaluation of the products that follow it. If the product is intensely disliked, the subsequent product will likely get a below-average evaluation. Also, this lingering effect diminishes proportionally as the person evaluates other products. Similarly, in pharmaceutical trials, while comparing more than two drugs under a crossover design, the carryover effect of a drug in a certain period can be significantly different if the same drug is administered in the next period compared to the carryover effect if a different drug is applied in the next period. This is because the impact of a drug is significantly dependent on the combination in which the other drugs are administered. \cite{senn2002cross} provides a mathematical explanation to justify that this simple carryover model is implausible in the pharmacokinetic study (see Chapter $10$ for details).

These problems can be addressed by including an interaction between treatment and carryover effects \citep{kershner1981two}. However, including an interaction term is feasible for a two-treatment design (d.f. for the interaction is 1), but for a study with several treatments, many interaction terms will be present in the model. To tackle this difficulty, \cite{afsarinejad2002repeated} introduced a model that differentiates self-carryover and mixed carryover effects. \cite{kempton2001optimal} introduced another model where the carryover effect of a treatment in the current period is proportional to the treatment effect in the previous period. Another approach is to include a second-order carryover term in the model. That means the response at period $j$ will also be affected by the treatment at period $j-2, j \geq 3$. An excellent repository of models with different carryover effects can be found in \cite{ozan2010assessing}. \rev{All the above models require the analyst to prespecify the form of carryover. To relax this, \cite{cruz2026penalized} recently proposed a penalized semiparametric generalized estimating equation (GEE) approach that estimates complex first-order carryover effects, which vary with the treatment order and the interaction between consecutive treatments, in crossover designs with repeated measurements.} We assume that all the parameters are estimable through within-subject differences. A general model is
\vspace{-0.1 cm}
\begin{align}
    Y_{ijk} = \mu + \pi_j + \alpha_i + \beta_{d(i,j)} + S_{ik} + \V{\gamma}^\top \M{B}_{ik} + \epsilon_{ijk} \label{eqn:GenModel}
\end{align}
where $\alpha_i$ is the effect of ith group, $\beta_{d(i,j)}$ is the effect of the treatment and carryover associated with $j$th period for $i$th group and \rev{$\M{B}_{ik}$} be the vector of baseline covariates measured for subject $k$ in group $i$. Any baseline covariates could also be included in the basic model~(\ref{trad_mod}), defined for AB-BA design. Depending on the design, one can include treatment by period interaction term and other factors in the model. Different forms of $\beta_{d(i,j)}$ are summarized in chapter 4 of \cite{jones2014design}.

\subsection{Optimal designs}
In a crossover study with $t$ treatments, we aim to efficiently estimate any pair of treatment contrast $\tau_i - \tau_j$. To accomplish that, we want to design the study in such a way that the SE of the estimator, $\Var(\hat{\tau}_i - \hat{\tau}_j)$ is minimized compared to that for any competing design. \cite{kiefer1975construction} introduced different optimality criteria based on the total variance of the estimator of all possible treatment contrasts. This is essentially a function of the trace of information matrix $C_d$ under the model assumed. Similarly, optimality criteria for an estimation of the carryover effect can also be defined.
\cite{hedayat1978repeated} pioneered finding optimal designs by considering a subclass of design called uniform designs. In a uniform design, each treatment is assigned to the same number of subjects in each period, and for each subject, every treatment is applied in the same number of periods. Uniformity is appealing because the treatment and carryover effect becomes orthogonal to each other, and it implies $p=\lambda_1t$ and $g=\lambda_2t$, where $\lambda_1, \lambda_2$ \rev{are positive integers}. \rev{Assuming} $p=t$ and $g=\lambda_2t$, they proved that balanced uniform (BU) designs are universally optimal for estimating treatment and carryover effects over the class of uniform designs. In a balanced design, each treatment is preceded by every other treatment an equal number of times, and a treatment can't be preceded by itself. Balance makes the standard error of all pairs of treatment contrasts the same.

\cite{cheng1980balanced} introduced a new class of design called strongly balanced designs and proved that any strongly balanced uniform design is universally optimal for estimating direct and carryover effects over the entire class of designs $d(t, p, g)$. A strongly balanced design is balanced with the relaxation that treatment can precede itself. If a design is strongly balanced, the number of groups (g) must be a multiple of $t^2$ and $p/t = m > 1$. An example of a two-treatment strongly balanced design is ABBA-BAAB-AABB-BBAA \citep{kershner1981two}. This means when $p > t$, these designs repeat the same treatment in consecutive periods. \cite{kunert1984optimality} worked with $p=t$ and \rev{showed} that if there exists a BU design, then the ratio of the trace of the information matrices of BU design and any competing design has a lower bound of $1 - (t^2 - t-1)^{-2}$. This means that BU designs are nearly optimal for $t>2$ but not strictly optimal. Kunert also proved that for $t=2=p$, the optimal designs for estimating direct treatment and carryover effects are different, and none is uniform.

 \cite{sen1987optimal} proved that when an interaction between treatment and carryover is present in the model, strongly balanced designs are still optimal for estimating direct treatment effects because they are orthogonal. \cite{kempton2001optimal} developed optimality results when carryover effects are proportional to treatment effects\rev{, which were further extended by \cite{bailey2006optimal}}. \cite{kunert2002optimal} developed optimal designs in the presence of self and mixed carryover effects by introducing a new class of designs called {\em totally balanced} designs. When the objective is to compare \rev{several} test treatments with a control treatment, \cite{hedayat2005optimal} obtained optimality results with $p=t+1$, called extra-period designs. Many works have been done to find optimal designs when $p < t$, we refer the reader to chapter 3 of \cite{bose2009optimal}. \rev{All of the above results are derived under a linear model for a continuous response. For non-Gaussian responses, \cite{jankar2020optimal} obtained locally D-optimal crossover designs under generalized linear models, where the parameters are estimated using GEE, and showed through simulations that the optimal allocations for two-treatment designs are reasonably robust to the choice of the working correlation structure. \cite{jankar2023general} subsequently established a general equivalence theorem for verifying the optimality of such designs.} We conclude our discussion on optimal designs by making two points: First, one should prioritize the effects or the contrasts that are of primary concern before choosing a design. Second, in the possibility of subject drop-out, will the design still be nearly optimal if the experimenter has to stop the study before the last period? \rev{A partial answer to the latter was provided by \cite{godolphin2019robustness}, who identified \textsl{perpetually connected} uniformly balanced designs, i.e., designs for which all treatment contrasts for direct and carryover effects remain estimable irrespective of the drop-out pattern, provided every subject completes the first two periods.} R-package \verb|Crossover| \citep{Crosspack} offers an excellent interface to choose between several designs under different carryover models by comparing the average efficiencies of the competing design under different correlation structures of error such as independence, compound symmetry (CS), etc.

\subsection{Inference methods for continuous response} \label{InfCont}
Depending on the design and the nature of experiments, the analyst might choose any model from~(\ref{eqn:GenModel}) for the treatment and carryover effects with some baseline covariates recorded at the start of the study. For now, we assume that a single response is observed within a period for every subject. As stated before, crossover design with $p$ periods can be considered split-plot experiments where the whole-plots are the subjects, and measurements are observed within each subject at different periods. Standard ANOVA $F$-test techniques for split-plots using random subject effect (whole plot error) can be conducted to test for treatment or carryover effects under a Gaussian assumption using \verb|proc mixed| or R package \verb|lmerTest| \citep{lmerTestPack}\rev{.}

Alternatively, crossover design can also be viewed as a repeated measurements study where measurements are observed for each subject over consecutive periods. Defining, $\M{Y}_{ik} = (Y_{i1k}, \cdots, Y_{ipk})^\top$ as the vector of response for the kth subject in sequence $i$, the random subject effect model in~(\ref{trad_mod}) induces an exchangeable or CS covariance structure for the response vector i.e. $\Var(\M{Y}_{ik}) = \M{\Sigma}= \sigma^2\M{I}_p + \sigma_s^2 \M{J}_p \forall \; i,k$. However, this CS structure might be too parsimonious to capture the true covariance. To generalize this we specify the mean model as $E(\M{y}_{ik} \; \vert \;\M{X} ) = \M{X}_{ik}\V{\theta}$. The fixed parameters $\V{\theta}$ contain all the fixed parameters specified in the mean model assumed, and the design matrix $\M{X}_{ik}$ is specified accordingly. Furthermore, the covariance model is $\Var(\M{y}_{ik} \; \vert \;\M{X}) = \M{V}_{ik}(\xi)$, which need not be compound symmetric. The vectors $\V{\theta}$ and $\V{\xi}$ are called mean and covariance parameters. This method directly captures the effect of treatment or carryover on the population mean response, which is called a population-averaged (PA) model. The influence of any random subject effect is captured in the complex covariance matrix. We assume that the responses for different subjects are mutually uncorrelated.

Under Gaussian assumption, the mean parameter $\V{\theta}$ is estimated through the generalized least square (GLS) approach and covariance parameters can be estimated using the restricted maximum likelihood (REML) method to adjust for the bias in the estimation of covariance. If normality is not assumed, then the mean component is estimated using linear estimating equations. SAS procedure \verb|proc mixed|, \verb|proc glimmix| and R-package \verb|nlme| \citep{nlmePack} \rev{offer} flexible PA models\rev{; the more recent R package \texttt{CrossCarry} \citep{crosscarry2023} implements GEE estimation specifically tailored to crossover data, with carryover terms and crossover-specific correlation structures}. The covariance structure can be varied according to any baseline covariate, e.g., the subject's sex. Many different covariance matrix structures can be assumed, such as Autoregressive (AR), CS, Ante, and Unstructured (Un), the most general one.  The final choice of the covariance model is made by AIC or BIC criterion. If the model is correctly specified, then GLS estimator of $\V{\theta}$ with its asymptotic distribution is given by
\begin{align}
    \hat{\V{\theta}} = \fpr{\sum_{i,k}\M{X}_{ik}^\top\M{V}_{ik}^{-1}(\hat{\xi})\M{X}_{ik}}^{-1}\sum_{i,k}\M{X}_{ik}^\top\M{V}_{ik}^{-1}(\hat{\xi}) \M{y}_{ik} \sim \textrm{Asymptotically N}\fpr{\V{\theta}, \hat{\M{\Sigma}}_M} \label{eqn: asymptheta}
\end{align}
where $\hat{\xi}$ is the REML/ML estimator of $\xi$ and $\hat{\M{\Sigma}}_M = \fpr{\sum_{i,k}\M{X}_{ik}^\top\M{V}_{ik}^{-1}(\hat{\xi})\M{X}_{ik}}^{-1}$ is called the model-based \rev{covariance matrix}. If the sample size in each group is not large,  adjustment due to \cite{kenward1997small} can be done. However, this adjustment still understates the SE of the estimator if the covariance structure is incorrectly specified. \cite{diggle2002analysis} pointed out in chapter 4 that the assumption of diagonal covariance does not affect the \rev{consistency of the} estimator of the mean parameter $\V{\theta}$. Still, it overstates or understates the SE of the estimator. The robust or empirical SE introduced in the context of non-normal response using a generalized estimating equation (GEE) gives protection against wrongly specified covariance \citep{zeger1988models}. The authors proved that under certain assumptions, as long as the mean model is correctly specified, the GLS estimator of the mean parameter still follows an asymptotic normal distribution with the covariance matrix replaced by \rev{the sandwich estimator} $\hat{\M{\Sigma}}_R$ \citep{yuan1998asymptotics}, which \rev{remains valid even when the working covariance is misspecified}. The empirical option in SAS \verb|proc mixed| allows inference on mean parameters using robust standard error. Adjustment to the empirical SE due to \cite{mancl2001covariance} can be done for a small sample. Wald-type test for testing $H_0: \M{C}\V{\theta}=\phi$ and approximate CI can be constructed using the test-statistic
\begin{align}
    T = \fpr{\M{C}\hat{\V{\theta}}-\phi}^\top\fpr{\M{C}\hat{\M{\Sigma}}\M{C}^\top}^{-1}\fpr{\M{C}\hat{\V{\theta}}-\phi} \sim \chi^2_r \quad \textrm{asymptotically} \;\; ; \quad \;\;  r=\textrm{rank}(\M{C}) \label{eqn: chisqGen}
\end{align}
If the primary questions of interest are to investigate the effect of treatment on each subject, then a subject-specific (SS) linear mixed effect model is appropriate. If $\V{b}$ is the vector of random effects then a SS model can be written as $E(\M{y}_{ik} \; \vert \;\M{X}, \V{b} ) = \M{X}_{ik}\V{\theta} + \M{Z}_{ik}\V{b}$. Usually, $\V{b}$ is assumed to follow Gaussian with mean $0$ and unknown covariance $\M{D}$. The PA model that generates from the SS model by taking an expectation over $\V{b}$ is $E(\M{y}_{ik} \; \vert \;\M{X} ) = \M{X}_{ik}\V{\theta}$, $\Var(\M{y}_{ik} \; \vert \;\M{X} ) = \M{Z}_{ik}\M{D}\M{Z}_{ik}^\top + \M{R}_{ik}$, where $\M{R}_{ik}$ is the covariance of the measurement error, possibly diagonal. This generates an equivalent PA model. However, in pharmacokinetic studies or when the carryover effect is assumed to be proportional to the treatment effect, the model becomes non-linear \citep{kempton2001optimal}. An SS model can be defined as $E(\M{y}_{ik} \; \vert \;\M{X} , \V{b} ) = \V{f}(\M{Z}_{ik}, \V{\theta}, \V{b})$ and an appropriate SS covariance model with parameter $\V{\psi}$. A PA model can be derived from this by integrating the likelihood over the distribution of $\V{b}$. Numerical approximations such as the adaptive quadrature method, a first-order approximation of likelihood are typically used to approximate the marginal distribution, and the estimation of $\V{\theta}$ and $\V{\xi}$ is carried out through the GEE method or maximizing the approximated likelihood directly. See \cite{davidian2003nonlinear} for a series of methodologies available in this context. SAS procedure \verb|proc nlmixed| implements this method with great efficiency.


\vspace{-0.2 cm}
\section{Using baseline measurements : AB-BA design revisited} \label{baseline}
The test for the carryover effect is based on subject totals and lacks power due to the high between-subject variability. Consequently, the PAR test for treatment effect becomes inefficient. \cite{hills1979two} proposed using baseline measurements collected before the start of each period in the study to potentially improve the power of the tests described in Section~\ref{AB-BA}. In many real applications, baseline measurements are collected only before the start of the first period. Using the baseline observations as responses, we can interpret the study as a four-period, two-treatment crossover trial. Moreover, it is intuitive to include an effect $(\theta)$ that differentiates between the two baseline responses due to the effect of the treatment in period-1, apart from the usual period ($\pi$), direct treatment $(\tau)$ and carryover effect $(\lambda)$.  Analysis can be done using the methods described in Section~\ref{InfCont} by specifying a suitable covariance structure for the four-variate response. In the presence of a significant $\lambda$ and $\theta$, the direct treatment effect is estimated using the difference between the first baseline and first period responses. If the model is specified correctly, this approach can provide a powerful test for the treatment effect as the between-subject variability is eliminated. The difference between the two baseline responses estimates $\theta$ \citep{patel1983use}. When both $\lambda$ and $\theta$ are null, this method produces the same within-subject estimator for $\tau$ as in~\rev{(\ref{eqn:treatEst})}. Non-parametric analysis can also be done following the rank-measure-based method by \cite{jung1999multivariate}.

While many people adopted this procedure to estimate treatment effect with more power, \cite{fleiss1985adjusting} presented a very critical review on using this method by showcasing that if $\theta$ behaves proportionally to the treatment effect $\tau$, then an induced or hypothetical carryover might appear even if it is genuinely insignificant. Subsequently, the inference drawn from the method can go wrong. \rev{\cite{willan1986baseline}} considered a random subject by period interaction and concluded that using baseline measurements can significantly diminish the power and precision if the subject by period variation is less than the within-subject variation, which is often the case in real-life applications, and this method should be applied with caution.

\rev{The debate on how best to use period-specific baselines has been largely settled in the last two decades. \cite{kenward2010use} and \cite{metcalfe2010analysis} revisited the analysis of crossover trials with period-specific baselines, treating the baselines as covariates in the model rather than as additional responses. \cite{mehrotra2014recommended} compared a range of approaches, including ignoring one or both baselines, change-from-baseline analysis, using functions of the baselines as period-specific or period-invariant covariates, and joint modeling of baseline and post-baseline responses. He showed that the power depends heavily on the structure of the within-subject covariance matrix and on how the baselines are used. He advised against the commonly used change-from-baseline analysis and recommended an analysis of covariance in which the within-subject difference in post-baseline responses is the outcome and the within-subject difference in baselines is the covariate. \cite{jemielita2016improved} further improved the power by using linear combinations of baselines as covariates, with the combination selected by an information criterion, while maintaining the type-I error rate, and \cite{jemielita2019efficient} extended these ideas to incomplete block crossover designs. Similar efficiency gains from period-specific baselines were demonstrated for censored time-to-event endpoints by \cite{xu2018incorporating}.}
\vspace{-0.4 cm}
\section{Analysis of Multivariate Continuous Response} \label{multivariate}
A multivariate response can arise in multiple practical applications. 1) The response that is of interest is inherently multivariate, and a single observation is recorded within each period. 2) The response might be univariate, but repeated measurements will be recorded within each period. 3) The response is multivariate and repeated measurements are observed within a period. \cite{wallenstein1979inclusion} considered a split-plot model when the response is single variate repeatedly measured over time within each period. The author considered a random effect structure for the subject by time interaction and analyzed the model using a univariate ANOVA approach. However, the model induces a CS structure for the response vector. \cite{patel1980multivariate} considered a common unstructured covariance for each response and analyzed through \rev{Wilks'} lambda test-statistic. \cite{Johnson1996thesis} in his Ph.D. thesis, developed a general multivariate ANOVA procedure to test for any effect in an AB-BA crossover design, considering that the multivariate response can occur from any of the three cases stated above. To give an overview of the approach, let $E(\M{Y}_{ijk}) = \V{\mu}_{ij}$ be the respective mean vector for the response vector at period $j$ of the group $i$. Defining the matrix of mean vectors $\mathcal{M} = \fpr{(\V{\mu}_{ij})}$, any hypothesis of fixed effects can be written as $H_0 : \M{C}\mathcal{M}\M{U} = \M{0}$, for appropriate coefficient matrices $\M{C}$ and $\M{U}$. The likelihood ratio test has been proposed to test for different fixed effects. The test statistics generally follow an F-distribution with appropriate degrees of freedom. \cite{johnson1993multivariate} proposed a non-parametric test using a multivariate generalization of Wilcoxon rank-sum statistic.

If a univariate response is recorded repeatedly over time within a period, then the ANOVA methods described above consider time as a factor and require the data to be balanced. To overcome this and to characterize the pattern of the response profile as a smooth function of time, a PA or an SS approach that models the response as a linear or polynomial function of time can be more appealing (chapter 5 of \citealt{jones2014design}). Inference for fixed effects can also be carried through the results described in Section~\ref{InfCont}. \rev{Later}, \cite{kawaguchi2010multivariate} developed a multivariate Mann-Whitney type non-parametric approach for a multivariate ordinal response following \cite{jung1999multivariate} under a two-period three-treatment crossover design. \rev{For repeated measurements within periods, \cite{cruz2024correlation} proposed a family of correlation structures for both Gaussian and non-Gaussian responses, in which the overall working correlation matrix of the GEE is the Kronecker product of a within-period and a between-period correlation matrix. This structure substantially reduces the number of correlation parameters compared to an unstructured matrix. \cite{cruz2023semiparametric} further proposed semiparametric GEE for repeated measurements within periods that allow the estimation of both simple and complex carryover effects, and \cite{sun2025dynamic} studied how the treatment effect evolves over the repeated measurements within a period. These methods are implemented in the R package \texttt{CrossCarry} \citep{crosscarry2023}.}
\vskip 0.1 pt
\section{Analysis of Binary/Categorical response} \label{binary}
In many real-life applications, the response can be categorical, i.e., the response can take only a few categories. See \cite{senn2002cross}, \cite{jones2014design} for example. Binary response occurs when the response can assume only two categories. This section will review the inference methodologies developed for binary and categorical responses under a crossover design. First, we start with the AB-BA design. We stick to the earlier notation for a typical response, but we say $Y_{ijk}=1$ if the response for $k$th subject in $i$th group in period $j$ is a \textsl{success} and $0$ if it is a \textsl{failure}. For example, suppose we want to compare the \rev{efficacy} of a drug (A) for cerebrovascular deficiency to a placebo (B). The response is the indicator variable denoting whether the cardiologist considers the electrocardiogram Normal (success) or abnormal (failure). Using the crossover difference $C_{ik}$ in~(\ref{eqn: notation}), define
\begin{align}
    N_i = \sum_{k=1}^{n_i} \mathbb{I}(\lvert C_{ik} \rvert =1) \; ; \quad
    N_i^A = \sum_{k=1}^{n_i} \mathbb{I}(C_{ik}=1) \; ; \quad
    N_i^B = \sum_{k=1}^{n_i} \mathbb{I}(C_{ik}=-1) \label{eqn: NonParPref}
\end{align}
Then, $N_i$ is the total number of subjects in sequence $i$ who show any \textsl{preference} to A or B. Out of them, $N_i^A$ showed a \textsl{preference} for A, and $N_i^B$ showed a \textsl{preference} for B. To illustrate the term \textsl{preference}, we say that a \textsl{preference} occurs if the responses for a subject under treatment A and B differ. Moreover, there is a \textsl{preference} towards treatment A if the response under treatment A is a \textsl{success} and that under treatment B is a \textsl{failure}, \rev{and} vice-versa for the interpretation of \textsl{preference} towards treatment B. For now, we assume that there is no carryover effect. Under the assumption that there is no period effect  ($\pi=0$), \cite{mcnemar1947note}'s test for paired responses can be adapted to test for treatment effects. \rev{This is because} under $H_0: \tau = 0$, $N_1^A + N_2^A$ follows binomial with parameter $N_1+N_2$ and $0.5$. Then, we can either do an exact or approximate $\chi^2$ test with one degree of freedom depending on the sample size.

McNemar's test does not account for the period effect, which is often significant in crossover designs. \cite{mainland1964elementary} proposed a heuristic test for treatment that adjusts for the period effect based on the association between \textsl{preference} to the treatments and the groups. If $\tau=0$, then the odds in favor of treatment A to B in group 1 should be the same as the odds in favor of treatment A in group 2. A chi-square test of independence or Fisher's exact test in a $2\times 2$ contingency table with one d.f can be constructed. \cite{gart1969exact} provides a theoretical justification of the test using a logistic model conditional on (fixed) subject effects. Mainland-Gart and McNemar tests discard the responses when \rev{$\lvert C_{ik} \rvert =0$}, i.e., the responses which show no \textsl{preference} to any treatment. \cite{prescott1981comparison} proposed a test which is similar to \rev{the} \cite{tudor1994review} approach described in \rev{Section}~\ref{AB-BANonPAr}. It considers a $2 \times 3$ contingency table where the extra \rev{column includes} the responses that show no \textsl{preferences}. This gives a chi-square test with $2$ degrees of freedom asymptotically. The simulation studies suggest that in the absence of the period effect, the McNemar and Prescott tests are more powerful than the Mainland-Gart test in detecting the treatment effect. The Prescott test performs better than both in the presence of a period effect.

All the methodologies discussed above were developed in terms of a contingency table and are useful for testing effects specifically under AB-BA design. Parametric marginal models have also been proposed for binary response under higher-order crossover designs. Unlike the PA model for a continuous response, the mean and covariance parameters are functionally related for a binary response. Following \cite{diggle2002analysis}, we write the population-averaged generalized linear model as follows
\vspace{-0.2 cm}
\begin{align}
    P(Y_{ijk}=1) = E(Y_{ijk}) = \V{f}\fpr{\V{X}_{ijk}, \V{\theta}} \quad \Var(Y_{ijk}) = \V{f}\fpr{\V{X}_{ijk}, \V{\theta}}(1-\V{f}\fpr{\V{X}_{ijk}, \V{\theta}}) = v_{ijk}(\V{\theta}) \label{eqn: BinaryPA}
\end{align}
where $\V{f}$ is an appropriate \rev{inverse link function} that maps onto $(0,1)$. In most \rev{cases}, $\V{f}\fpr{\V{X}_{ijk},  \V{\theta}} = 1/(1 + \exp(-\V{X}_{ijk}^\top \V{\theta}))$. However, the specification of the marginal variance is insufficient to capture the correlation between the two responses for a subject. Also, for binary responses, the parameter space of the correlation is restricted. In the literature, it is standard to assume a working correlation structure. Assuming a working correlation matrix $\M{\Gamma}(\V{\alpha}, \V{X}_{ik})$ for $\M{Y}_{ik}$, we get the marginal covariance matrix as $\Var(\M{Y}_{ik})= \M{V}_{ik}(\V{\theta}, \V{\alpha}) = \tpr{\textrm{Diag}((v_{ijk}))}^{1/2}\Gamma(\alpha, \V{X}_{ik})\tpr{\textrm{Diag}((v_{ijk}))}^{1/2}$. Consistent estimation of $\V{\theta}$ and $\V{\alpha}$ can be done by GEE \citep{liang1986longitudinal} or quadratic estimating equation (GEE-2, \citealt{prentice1991estimating}). Under certain technical assumptions, the large sample distribution of $\V{\theta}$ has the same form as we obtained in~(\ref{eqn: asymptheta}), provided the mean model is correctly specified. In practice, the covariance structure is \rev{often} incorrectly specified, and a robust covariance matrix is used to draw inferences about $\V{\theta}$. This method is implemented in SAS procedure \verb|proc genmod, proc gee, proc glimmix| and \verb|geepack| in R \rev{\citep{geepack1}}. \rev{Optimal designs for such marginal models with binary and other non-Gaussian responses are discussed in \cite{jankar2020optimal}, and the GEE framework with crossover-specific correlation structures for non-Gaussian responses is available in \cite{cruz2024correlation}.}

Similarly, allowing for a subject-specific random effect, a generalized linear mixed model (GLMM) for binary response can be defined as follows,
\vspace{-0.2 cm}
\begin{align}
    E(\M{Y}_{ik}\vert \M{S}_{ik}) = \V{f}\fpr{\V{X}_{ik}, \V{\theta}, \M{S}_{ik}} \quad \Var(\M{Y}_{ik}\vert \M{S}_{ik}) = \M{R}_{ik}(\V{\theta}, \V{\alpha}, \V{X}_{ik}, \M{S}_{ik}) \label{eqn: SSBinary}
\end{align}
\vspace{-0.1 cm}
In general, it is assumed that given $\M{S}_{ik}$, the observations in each period are independently distributed so that $\M{R}_{ik}$ be diagonal. Then, the unconditional mean and covariance model for $\M{Y}_{ik}$ can be obtained by integrating over the distribution of $\M{S}_{ik}$, which is also usually assumed to be Gaussian. Large sample inference for $\V{\theta}$ can be carried out through first-order approximation, GEE, or Bayesian methods, implemented in SAS procedure \verb|proc glimmix|. See \cite{davidian2003nonlinear} for details.

If the response assumes M $(>2)$ categories, then the response can be transformed to $\M{Y}_{ijk} = \V{e}_m$ if the response is in the $m$th category\rev{, where} $\V{e}_m \in \Real^M$ \rev{is} the unit vector, $m=1,\dots, M$. Let $\pi_{ijk,m} = P(\M{Y}_{ijk}=\V{e}_m)$. Analogous to the extension of binary logistic regression to multinomial logistic regression, generalized logit model $\pi_{ijk,m}/\pi_{ijk,1} = \V{f}\fpr{\V{X}_{ik}, \V{\theta}_m} \;\; m=2,\dots,M$.
SAS procedure \verb|proc genmod| can fit a population average logit model for categorical response. If the response is ordinal, then the proportional odds model can also be fit that characterizes $P(Y_{ijk} \geq m), m=1,\dots, M$ through suitable link function and fixed effect parameters. See \cite{jones2014design} for further details.
\vspace{-0.15 cm}
\vskip 0 pt
\section{Conclusion and Open problems} \label{open}
Statistical methods developed for longitudinal study \citep{diggle2002analysis} can also be applied to any crossover design. In the AB-BA design, the carryover causes a nuisance to the efficient estimation of direct treatment. Including a sufficiently long washout period can be effective in this situation\rev{, although it cannot remove behavioral carryover \citep{shi2024behavioral}}. No particular procedure discussed in Section~\ref{AB-BAPar} outperforms the \rev{others} in all scenarios. We shortlist some open problems in the literature. Firstly, as a crossover study with more than two periods lasts long or multiple treatments are given on the subject in different periods, subject drop-out is inevitable. \rev{Likelihood-based analyses under the missing at random (MAR) assumption \citep{rosenkranz2015analysis} and models for nonignorable dropout \citep{wang2021nonignorable} have been proposed, and designs robust to drop-out have been identified \citep{godolphin2019robustness}. Moreover, the estimand framework of the ICH E9(R1) addendum \citep{ich2019e9r1} has made it clear that the treatment of drop-out determines the target of inference. For instance, \cite{matthews2023contrasting} showed that a complete-case analysis of a complete crossover design estimates a principal stratum estimand (the effect among patients who can tolerate all treatments), which in general differs from the hypothetical estimand that would be obtained if all patients completed the trial. Nevertheless,} a significant amount of work is still to be done when the \rev{MAR} assumption is invalidated\rev{, and on choosing and estimating clinically relevant estimands in multi-period crossover trials}. Secondly, \rev{although GEE-based and semiparametric methods for repeated measurements within a period have recently been developed \citep{cruz2024correlation, cruz2023semiparametric, sun2025dynamic}, a functional data analysis framework that treats the within-period response profile as a smooth random function, with} non-parametric or semi-parametric estimation of \rev{the functional treatment and carryover effects} along with \rev{their} asymptotic properties, is yet to be explored. Thirdly, from a design perspective, if the carryover is proportional to direct treatment, then the optimal design developed assumes that the proportionality parameter is known. The construction of a robust design for the correct estimation of the unknown parameter might be essential. \rev{Similarly, the locally optimal designs for generalized linear models \citep{jankar2020optimal} depend on the unknown parameter values, and robust or Bayesian alternatives deserve further study.} Lastly, one of the primary advantages of the crossover design is that it requires a comparably small number of subjects in the analysis. However, the small sample properties of the estimators we discussed are still to be examined rigorously\rev{, particularly for the recently developed covariate-adjusted \citep{shi2024behavioral} and GEE-based \citep{cruz2024correlation} estimators, which rely on asymptotic approximations}.


\bibliographystyle{agsm}
\bibliography{ref.bib}

\end{document}